\documentclass[10pt,letterpaper]{article}
\usepackage{opex3}
\usepackage{amsmath}

\begin{document}

\title{Bichromatic homodyne detection of broadband quadrature squeezing}

\author{Christopher S. Embrey$^1$, Joshua Hordell$^1$, Plamen G. Petrov$^{1,*}$, and Vincent Boyer$^1$}

\address{$^1$Midlands Ultracold Atom Research Centre, School of Physics and Astronomy, University of Birmingham, Edgbaston, Birmingham B15 2TT, UK}

\email{$^*$p.g.petrov@bham.ac.uk} 



\begin{abstract}
We experimentally study a homodyne detection technique for the characterization of a quadrature squeezed field where the correlated bands, here created by four-wave mixing in a hot atomic vapor, are separated by a large frequency gap of more than 6~GHz. The technique uses a two-frequency local oscillator to detect the fluctuations of the correlated bands at a frequency accessible to the detection electronics. Working at low detection frequency, the method allows for the determination of both the amplitude and the phase of the squeezing spectrum. In particular, we show that the quadrature squeezing created by our four-wave mixing process displays a noise ellipse rotation of $\pi/2$ across the squeezing spectrum. 
\end{abstract}






\section{Introduction}\label{sec:introduction}
Squeezed light, where the uncertainty on one of the field quadratures is brought below the limit given by the balanced application of the Heisenberg uncertainty principle, can improve the sensitivity of classical measurements when their operation is limited by the quantum noise of classical light. It has been experimentally demonstrated that squeezed light can increase the signal-to-noise ratio in spectroscopy \cite{polzik}, theoretically proposed to improve the sensitivity of interferometry in large band of frequencies \cite{unruh}, and it has been suggested that spatially multimode squeezed light can improve the resolution of optical imaging \cite{kolobov}. In quantum information science squeezed states are the workhorse for obtaining continuous variable entanglement for unconditional protocols \cite{braunstein}. The state-of-the-art methods for the generation of squeezed light include nonlinear processes parametric down-conversion (PDC) in solid-state crystals \cite{wu}, four-wave-mixing (4WM) in atomic vapours \cite{slusher} and both PDC and 4WM in optical fibres \cite{shelby}. 

In the case of continuous pumping, these nonlinear parametric processes induce quantum correlations over a finite bandwidth on pairs of sidebands symmetrically placed about a center frequency (CF), which is determined by the frequency of the pump light. If the nonlinear process is frequency degenerate this results in a continuous range of pairwise correlated frequency sidebands centered on and containing the CF. If the process is frequency non-degenerate the correlated sidebands are contained in two disjointed frequency ranges symmetrically placed with respect to the CF. The two possible superpositions of two correlated sidebands both display squeezing on one of their quadratures and they are collectively referred to as a two-mode squeezed state (TMSS).

The orthodox scheme to detect quadrature squeezing is the balanced homodyne detection \cite{yurke}, whereby the correlated sidebands beat with a strong local oscillator (LO) field at the CF. This method is most appropriate in the degenerate case, where the frequencies of the LO and the sidebands are close. In this case the squeezing spectrum, i.e. the range of frequencies at which the beats display quantum correlations, should start at DC. Noise measurements near DC however are greatly influenced by 1/f technical noise, so in practice the light noise is measured at a high enough analysis frequency (AF), away from the technical noise band.

In the non-degenerate case, where the frequency separation between correlated sidebands are potentially large, the homodyne detection of these states with a LO at the CF can become experimentally challenging due to the limited frequency response of the low-noise photodetectors and electronics. The solution to this problem is to use a two-frequency-component local oscillator, i.e. a bichromatic local oscillator (BLO), also referred to as two-tone LO. The BLO homodyne detection method has been theoretically proposed for configurations where the correlated sidebands are separated by the use of a cavity \cite{zhang} and for arbitrary sideband frequency separations by Marino et al. \cite{marino}. Since then there have been few realizations of the concept. For instance multi-frequency LO homodyne detector set-ups have been experimentally realized for the detection of entanglement in cluster states of optical frequency combs in both CW \cite{pysher} and pulsed \cite{roslund} regimes. More relevant to the present work, quadrature-squeezed light generated in optical parametric oscillators (OPO) has been detected using BLO employing frequency separation of few MHz, in order to avoid limitations imposed by technical noise at low analyzing frequencies \cite{li}. In atomic vapors, the BLO technique has been recently used to detect multi-spatial-mode quadrature squeezing in a continuous squeezing bandwidth with large sideband separation \cite{embrey}.

\begin{figure}[t!]
\centering
\includegraphics[width=12cm]{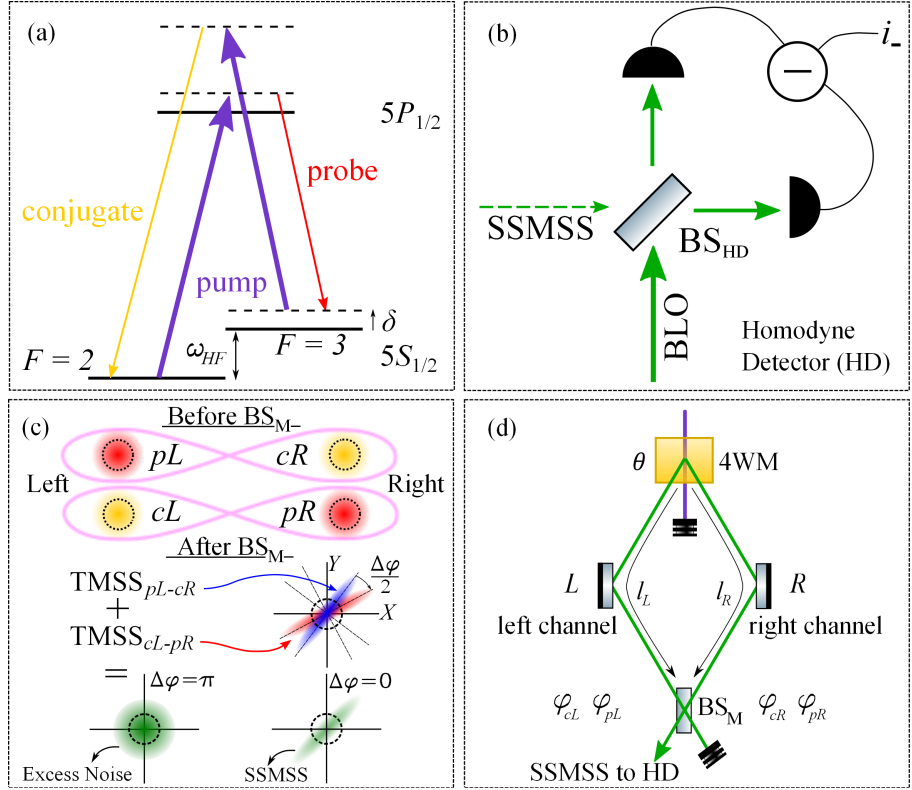}\\
\caption{Generation and detection of SSMSS state. In (a) is the level diagram of the 4WM process. In (b), the homodyne detection scheme, which can measure any quadrature of a squeezed state through interference with a LO on the $\mathrm{BS_{HD}}$. In this case the state is a SSMSS and is analyzed by a BLO. In (c), a diagram of the SSMSS generation via overlap of two TMSS on the $\mathrm{BS_M}$. Before the $\mathrm{BS_M}$, the light field is the superposition of two TMSS $pL-cR$ and $cL-pR$ which exist across the Left and Right spatial modes. The resulting SSMSS after the $\mathrm{BS_M}$, are mismatched in phase space by an angle $\Delta\varphi/2$. Their superposition is a true SSMSS for $\Delta\varphi = 0$. Note that for a value of $\Delta\varphi = \pi$ the resulting state exhibit excess noise on all quadratures. Everywhere the dotted line represents the noise of the vacuum. In (d), the bichromatic interferometer beam paths together with the corresponding phases and channels as described in the text. The letters $L$ and $R$ in (c) and (d) denote the Left and Right propagation modes, respectively.}
\label{fig:detection}
\end{figure}

In our experiment we use non-degenerate 4WM to create a TMSS that contains two correlated light modes separated by 6~GHz as shown in Fig.\ref{fig:detection}(a). As these modes are spatially distinct we combine them to generate a single spatial mode squeezed state (SSMSS). We show that the BLO homodyne detection setup can be used to fully reconstruct the squeezing spectrum of the SSMSS state. By controlling the phase of the BLO, we measure the rotation of the noise ellipse phase (NEP) which results from the dispersion of the medium. 

\section{Bichromatic homodyne detection}\label{sec:theory}
\subsection{Homodyne detection}\label{sec:HD}
The canonical technique to detect small fluctuations on a signal field is the balanced homodyne detection. This technique uses a LO that is overlapped with the signal mode to be analyzed on a 50/50 beamsplitter ($\mathrm{BS_{HD}}$). The two output intensities are detected by a balanced photodetector which forms the photocurrent difference [Fig.\ref{fig:detection}(b)] and the trans-amplified signal is spectrum analyzed. Let us call $\hat{d}_1$ and $\hat{d}_2$ the photon number operators in the two output arms of the $\mathrm{BS_{HD}}$. They can be expressed through the positive frequency part of the signal electric field $\hat{E}_S$ and the LO electric field $\hat{E}_{LO}$ as:
\begin{eqnarray}
\centering
\hat{d}_1 &=& \frac{1}{\sqrt{2}}\left(\hat{E}_{S} + i\hat{E}_{LO}\right) \nonumber \\
\hat{d}_2 &=& \frac{1}{\sqrt{2}}\left(i\hat{E}_{S} + \hat{E}_{LO}\right),
\label{eq:output_op}
\end{eqnarray}
up to a constant factor. With the above, the subtracted photocurrent has the form:
\begin{eqnarray}
\centering
\hat{i}_{-} &=& \hat{d}_1^{\dagger}\hat{d}_1 - \hat{d}_2^{\dagger}\hat{d}_2 = i\left(\hat{E}_{S}^{\dagger}\hat{E}_{LO} - \hat{E}_{LO}^{\dagger}\hat{E}_{S}\right).
\label{eq:photocurrent_op}
\end{eqnarray}
Let us consider a coherent state LO $\hat{E}_{LO} = \hat{b}e^{-i\omega_{LO}t}$ with $\langle\hat{b}\rangle = |\beta|e^{i\varphi_{LO}}$, where $\hat{b}$, $\beta$, $\omega_{LO}$, and $\varphi_{LO}$ are the annihilation operator, amplitude, frequency and phase at the $\mathrm{BS_{HD}}$ of the LO, respectively. The signal field contributing to the component of $\hat{i}_{-}$ at frequency $\Omega$ is then the two-mode field, up to a constant factor: 
\begin{equation}
\centering
\hat{E}_S = \hat{a}_{+}e^{-i(\omega_{+}t-\varphi_{LO})} + \hat{a}_{-}e^{-i(\omega_{-}t-\varphi_{LO})},
\label{eq:E_s}
\end{equation}
with $\omega_{\pm} = \omega_{LO}\pm\Omega$, and $\hat{a}_\pm$ are the annihilation operators of the corresponding signal bands at frequencies $\omega_{\pm}$. 

The quadratures of the signal field are
\begin{eqnarray}
\centering
\hat{X} &=& \frac{1}{2\sqrt{2}}\left(\hat{a}_{+} + \hat{a}^{\dagger}_{+} + \hat{a}_{-} + \hat{a}^{\dagger}_{-}\right) \nonumber \\
\hat{Y} &=& \frac{1}{i2\sqrt{2}}\left(\hat{a}_{+} - \hat{a}^{\dagger}_{+} + \hat{a}_{-} - \hat{a}^{\dagger}_{-}\right).
\label{eq:quadratures}
\end{eqnarray}
When the signal is a TMSS, correlations between the bands at frequencies $\omega_{+}$ and $\omega_{-}$ lead to fluctuations of the quadratures taking the form \cite{loudon}:
\begin{eqnarray}
\centering
\langle(\Delta\hat{X})^2\rangle &=& \frac{1}{4}\left(e^{-2s}\cos^2\frac{\theta}{2} + e^{2s}\sin^2\frac{\theta}{2}\right) \nonumber \\
\langle(\Delta\hat{Y})^2\rangle &=& \frac{1}{4}\left(e^{-2s}\sin^2\frac{\theta}{2} + e^{2s}\cos^2\frac{\theta}{2}\right),
\label{eq:quadratures_noise}
\end{eqnarray}
where $s$ is the degree of squeezing and $\theta$ is the squeezing angle, which is referenced to $\varphi_{LO}$. One can derive the variance of the homodyne detector signal in the limit of a LO much stronger than the signal \cite{loudon}:
\begin{equation}
\centering
\langle(\Delta\hat{i}_{-})^2\rangle = 2|\beta|^2\left[e^{-2s}\cos^2\left(\chi-\frac{\theta}{2}\right) + e^{2s}\sin^2\left(\chi-\frac{\theta}{2}\right)\right].
\label{eq:HD_signal_Mono_LO}
\end{equation}
We have introduced the phase $\chi$, which also accounts for the more general case where the phase reference for $\theta$ is taken at a different location from the $\mathrm{BS_{HD}}$. One can take $\chi = \chi_+ + \chi_- =\frac{1}{2}(\varphi_{LO} - \varphi_{+})+\frac{1}{2}(\varphi_{LO} - \varphi_{-})$, where $\varphi_\pm$ are the propagation phases for modes $\hat{a}_\pm$, and $\chi_{\pm}$ are the differences between the phase of LO and signal bands. The minimum noise is observed for $\chi=\theta/2$ and it is below the shot noise given by $2|\beta|^2$.

In the single frequency LO scheme, the signal is at the beat frequency $\Omega = \frac{\omega_{+}-\omega_{-}}{2}$ between the sidebands and the LO. This technique is more useful when the frequencies of the correlated sidebands are close i.e. for small $\Omega$, as analyzing quantum noise at high frequencies can be challenging or even impossible. For large $\Omega$, the solution is to use a bichromatic local oscillator with two frequency components near the corresponding frequencies of the TMSS correlated sidebands i.e. $\hat{E}_{BLO} = \hat{b}_{1}e^{-i\omega_{L1}t}+\hat{b}_{2}e^{-i\omega_{L2}t}$, with $\hat{b}_{1,2}$ the annihilation operators for the fields at frequencies $\omega_{L1,L2}$, where these components can be described as coherent states $\langle\hat{b}_1\rangle = |\beta_1|e^{i\varphi_{L1}}$ and $\langle\hat{b}_2\rangle = |\beta_2|e^{i\varphi_{L2}}$, with amplitudes $\beta_{1,2}$ and phases $\varphi_{L1,L2}$. The noise response at low analyzing frequency is then dominated by the beat of the correlated sidebands with their nearest BLO components. The theory of the BLO detection method is described in detail in \cite{marino} and here we will only write up the result for the subtracted photocurrent operator for the case when the LO components have the same amplitude and are symmetrically placed about the CF, i.e. for $\omega_{L1}-\omega_{-} = - (\omega_{L2} - \omega_{+})$ and $|\beta_1|=|\beta_2|=|\beta|$:
\begin{equation}
\centering
\langle(\Delta\hat{i}_{-})^2\rangle = 4|\beta|^2\left[e^{2s}\cos^2\left(\frac{\chi_1+\chi_2-\theta}{2}\right) + e^{-2s}\sin^2\left(\frac{\chi_1+\chi_2- \theta}{2}\right)\right].
\label{eq:HD_signal_BLO}
\end{equation}
The last equation shows that the homodyne detector signal depends on the relative phase differences $\chi_1 = \varphi_{L1}-\varphi_{-}$ and $\chi_2 = \varphi_{L2}-\varphi_{+}$ between the local oscillator components and their corresponding TMSS sidebands. Although Eq.~(\ref{eq:HD_signal_BLO}) was derived for a single pair of correlated sidebands, each frequency component of the BLO each interacts with a pair of sidebands therefore there exist two pairs of detected sidebands, lying within the squeezing spectrum or not, whose noises add in quadrature. The effect of these so-called image bands \cite{marino} is discussed later.

\subsection{Mode-matching of correlated sidebands in non-degenerate 4WM}\label{sec:overlap}
We consider non-degenerate 4WM in a hot Rb vapor. In this process two photons from a strong pump beam interact with a Rb atom to create a pair of probe (p) and conjugate (c) (or stokes and anti-stokes) photons at frequencies oppositely detuned from the pump as shown in Fig.\ref{fig:detection}(a). The TMSS state generated in this manner is entangled across probe and conjugate sidebands of two spatially distinct modes \cite{boyer}, separated by twice the hyperfine splitting of the ground state, $2\omega_{HF}$. The phase-matching condition of the 4WM process requires the generated TMSS components to propagate symmetrically about the pump axis at a finite angle \cite{turnbull}. In order to create the SSMSS, we overlap the two separated modes on a 50/50 beamsplitter ($\mathrm{BS_M}$, M for "mixing"). The result after the overlap is shown in Fig.\ref{fig:detection}(c). It is important to note that each of the left ($L$) and right ($R$) input modes of the $\mathrm{BS_M}$ contains probe and conjugate components [Fig.\ref{fig:detection}(c)] and that these components may interfere at the $\mathrm{BS_M}$ with different phases. This phase mismatch between the probe and conjugate components of the $L$ and $R$ channels plays an important role in the transformation from TMSS to SSMSS.

Let us consider the SSMSS state on one of the outputs of the $\mathrm{BS_M}$ as shown in Fig.\ref{fig:detection}(d). The operators for the fields at the probe and conjugate frequencies are:
\begin{eqnarray}
\centering
\hat{a}_{p} &=& \frac{1}{\sqrt{2}}\left(\hat{a}_{pL}e^{i\varphi_{pL}} + \hat{a}_{pR}e^{i\varphi_{pR}}\right) \nonumber\\
\hat{a}_{c} &=& \frac{1}{\sqrt{2}}\left(\hat{a}_{cL}e^{i\varphi_{cL}} + \hat{a}_{cR}e^{i\varphi_{cR}}\right),
\label{eq:operators}
\end{eqnarray}
where $\hat{a}_{mn}$ with $m = p,c$ and $n = L,R$ are the photon annihilation operators of probe and conjugate components in the $L$ and $R$ input channels and $\varphi_{mn}$ are their phases at the $\mathrm{BS_{M}}$. In other words the output is the superposition of two TMSS, containing probe in $L$ and conjugate in $R$ ($pL-cR$) and vice-versa ($cL-pR$). For symmetry reasons these two TMSS are described by a single complex squeezing parameter $s e^{i\theta}$. Substituting Eq.(\ref{eq:operators}) into Eq.(\ref{eq:quadratures}), replacing the indexes $-$ and $+$ with $p$ and $c$, we obtain the variance of the quadrature operators as \cite{loudon}
\begin{eqnarray}
\begin{aligned}
\langle(\Delta\hat{X})^2\rangle &=& \frac{1}{4}\biggl\{\frac{e^{2s}}{2}\left[1-\cos\left(\frac{\varphi_{t}}{2} + \theta\right)\cos\frac{\Delta\varphi}{2}\right]\\
&+&\frac{e^{-2s}}{2}\left[1+\cos\left(\frac{\varphi_{t}}{2} + \theta\right)\cos\frac{\Delta\varphi}{2}\right]\biggr\}\\
\langle(\Delta\hat{Y})^2\rangle &=& \frac{1}{4}\biggl\{\frac{e^{2s}}{2}\left[1+\cos\left(\frac{\varphi_{t}}{2} + \theta\right)\cos\frac{\Delta\varphi}{2}\right]\\
&+& \frac{e^{-2s}}{2}\left[1-\cos\left(\frac{\varphi_{t}}{2} + \theta\right)\cos\frac{\Delta\varphi}{2}\right]\biggr\},
\end{aligned}
\label{eq:variances}
\end{eqnarray}
where $\varphi_{t} = (\varphi_{pR} + \varphi_{pL}) + (\varphi_{cR} + \varphi_{cL})$ and $\Delta\varphi = (\varphi_{pR} - \varphi_{pL}) - (\varphi_{cR} - \varphi_{cL})$. The propagation phases $\varphi_t$ and $\delta\varphi$ can be expressed by the sum and difference of the geometrical paths $l_L$ and $l_R$ of left and right modes, and the probe and conjugate frequencies $\omega_{p}$ and $\omega_{c}$ as:
\begin{eqnarray}
\Delta\varphi & = & \frac{1}{c}(\omega_{p} - \omega_{c})(l_{R}-l_{L})\nonumber\\
	\varphi_{t} & = & \frac{1}{c}(\omega_{p} + \omega_{c})(l_{R}+l_{L}).
\label{eq:phases}	
\end{eqnarray}

From Eq.(\ref{eq:variances}) one can see that for $\Delta\varphi = \pi$ the $\hat{X}$ and $\hat{Y}$ quadratures of the combined state display noise above shot noise regardless of the squeezing angle $\theta$. On the contrary, the condition $\Delta\varphi = 0$ leads to variances similar to those in Eq.(\ref{eq:quadratures_noise}). This can be understood by looking at the overlap of the two TMSS at the $\mathrm{BS_{M}}$ in phase space [Fig.\ref{fig:detection}(c)]. Each component of the TMSS state ($pL-cR$ or $cL-pR$) is independently transformed into an SSMSS as shown in Fig.\ref{fig:detection}(c) and the best overall squeezing is obtained for $\Delta\varphi = 0$. This can be achieved by insuring that the difference between $l_{R}$ and $l_{L}$ is much smaller than the wavelength associated with the frequency difference $\omega_{p}-\omega_{c}$ [see Fig.\ref{fig:detection}(d)]. With these assumptions the noise in the quadratures is simplified to:
\begin{eqnarray}
\centering
\langle(\Delta\hat{X})^2\rangle &=& \frac{1}{4}\left[e^{2s}\sin^2\left(\frac{\varphi_{t}}{4} + \frac{\theta}{2}\right) + e^{-2s}\cos^2\left(\frac{\varphi_{t}}{4} + \frac{\theta}{2}\right)\right] \nonumber\\
\langle(\Delta\hat{Y})^2\rangle &=& \frac{1}{4}\left[e^{2s}\cos^2\left(\frac{\varphi_{t}}{4} + \frac{\theta}{2}\right) + e^{-2s}\sin^2\left(\frac{\varphi_{t}}{4} + \frac{\theta}{2}\right)\right].
\label{eq:variances_simple}
\end{eqnarray}
The last expression recovers the familiar form of the quadratures noise as given in Eq.(\ref{eq:quadratures_noise}) apart from a common phase $\varphi_{t}$. This extended analysis is necessary due to the bichromatic nature of the generated correlations\cite{footnote}.

The last step in the analysis is to obtain an expression for the SSMSS noise, as discussed in Sec.\ref{sec:HD}. In our case the correlations are bichromatic and the homodyne detector uses a BLO. Then the variance of the detector subtracted photocurrent is given by Eq.(\ref{eq:HD_signal_BLO}) and has the form
\begin{equation}
\centering
\langle(\Delta\hat{i}_{-})^2\rangle = 4|\beta|^2\left[e^{2s}\sin^2\left(\frac{\varphi^{'} - \theta}{2}\right)+ e^{-2s}\cos^2\left(\frac{\varphi^{'} - \theta}{2}\right)\right]
\label{eq:photocurrent}
\end{equation}
In the last expression we have substituted $\varphi^{'} = \chi_{p} + \chi_{c} $, with $\chi_{p} = \varphi_{p}^{LO} - \varphi_{p}$ and $\chi_c = \varphi_{c}^{LO} - \varphi_{c}$ being the phase differences between each of the components of the BLO, and the corresponding sideband of the SSMSS. 

\begin{figure}[t]
\centering
\includegraphics[width=10cm]{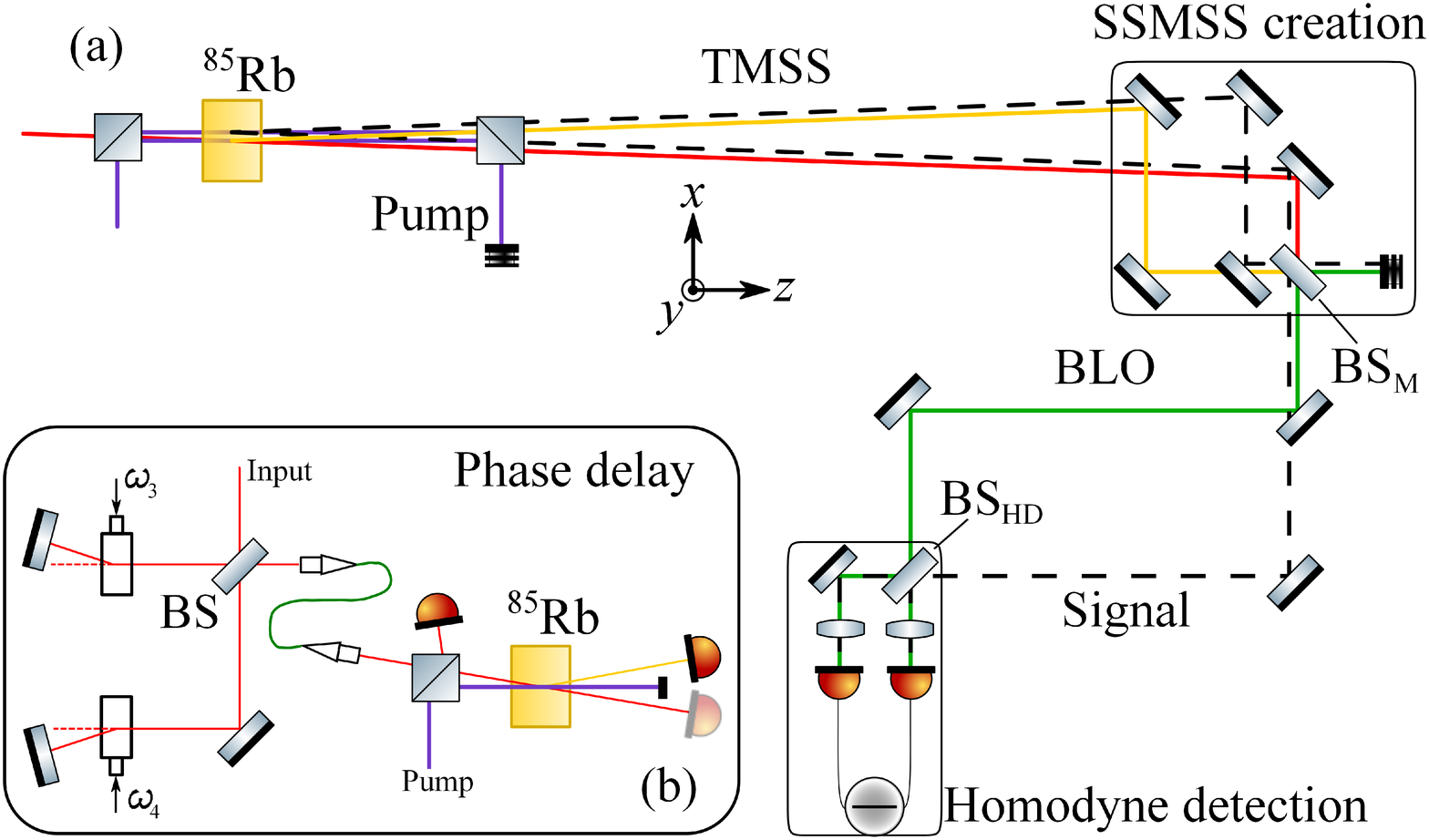}\\
\caption{Schematic diagram of the setup (modified from \cite{embrey}) . In (a), the creation and measurement of the SSMSS. The vacuum fields are depicted by black dashed lines and at all points contain the probe and conjugate frequencies. The bright fields propagation path is drawn in red and yellow representing probe and conjugate LO frequencies respectively. The green lines represent the combined probe and conjugate fields in a single BLO beam. The purple lines depict the pump light propagation path. The slight offset between the vacuum and LO fields in the diagram represents their vertical separation in the experiment. However, we use a single $\mathrm{BS_{M}}$ for both of them in the overlapping stage. In (b), simplified setup for phase delay measurements as described in Sec.\ref{sec:NEP_rotate}. The radio-frequency drives for the AOM's satisfy the condition $\omega_{4}-\omega_{3} = \delta/2$, with $\omega_{3}$ fixed close to $\omega_{HF}/2$.}
\label{fig:setup}
\end{figure}

\section{Experimental setup}
A simplified experimental setup is shown in Fig.\ref{fig:setup}(a). A heated rubidium cell is pumped by two pump beams separated vertically, producing two non-overlapping 4WM amplifiers. One of the amplifiers is seeded at $\omega_{p}$ to produce the bright components of the BLO at $\omega_{p}$ and $\omega_{c}$, while the other amplifier generates the correlated modes of the TMSS. The seed beam is generated via an acousto-optic modulator (AOM) in double-pass configuration with an RF drive frequency approximately matching half of the ground state hyperfine splitting $\omega_{HF}$.  A mixing beamsplitter, the same as $\mathrm{BS_M}$ in Sec.\ref{sec:overlap}, overlaps the modes of the TMSS and produces the SSMSS at one of the outputs. Equivalently, the bright fields are mixed on the same $\mathrm{BS_M}$ to produce the BLO. This method of producing the BLO ensures phase stability of the components of the BLO with respect to the signal as well as automatic spatial mode matching \cite{boyer,embrey}.

For detection, the BLO and the SSMSS from one of the $\mathrm{BS_M}$ outputs are further overlapped on a second 50/50 beamsplitter, the same as $\mathrm{BS_{HD}}$ in Sec.\ref{sec:HD}, at the balanced homodyne detection stage. The relative phase between the SSMSS and the BLO controls the measured quadrature and is tuned by changing the BLO pathlength with a piezo-electric actuator. This means that both $\varphi_{p}^{LO}$ and $\varphi_{c}^{LO}$ are scanned simultaneously. The output channels of the $\mathrm{BS_{HD}}$ are detected by a balanced photodetector and the produced photocurrent noise is observed on a spectrum analyser in zero span. Sample noise traces are presented in Fig.\ref{fig:homodyne_signal}. 

As described in Sec.\ref{sec:overlap}, adjusting the difference between paths $l_R$ and $l_L$ allows for efficient transformation of the TMSS correlations into a SSMSS. This is done by canceling the path difference between the gain region and the $\mathrm{BS_{M}}$, as shown in Fig.\ref{fig:detection}(d). Practically, a way to achieve this is to temporarily seed the 4WM process responsible for the squeezed vacuum generation at the probe frequency, in both the $L$ and $R$ directions. The two output modes $L$ and $R$ contain bright beams at probe and conjugate frequencies and form a bichromatic interferometer whose fringe visibility can be used as a benchmark for minimizing the pathlength difference $\Delta l = l_R-l_L$ ("white light" fringes). In the experiment, the pathlength difference is altered and the quality of the overlap is monitored by observing the interference contrast for the bichromatic light. The achieved visibility is more than 99$\%$, which indicates a pathlength difference of less than $2\mathrm{mm}$. 

\section{Results}
\subsection{Squeezing spectrum}\label{sec:spectrum}
\begin{figure}[t]
\centering
\includegraphics[width=8cm]{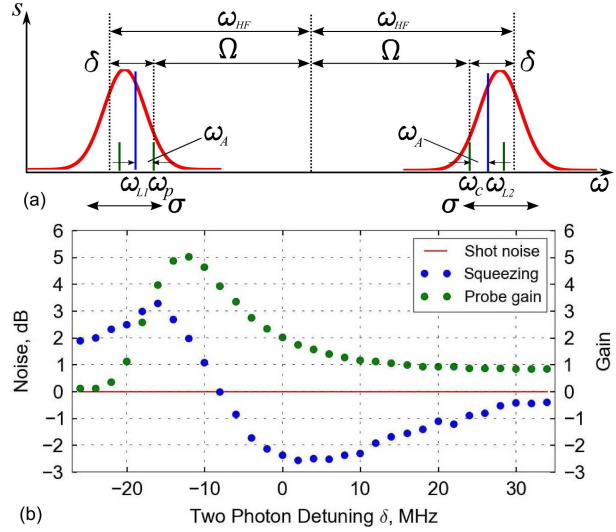}
\caption{(a) Correlations in frequency domain, $\omega_p$ and $\omega_{c}$ are the probe and conjugate sidebands frequencies and $\sigma$ is the squeezing bandwidth. Image bands shown in green, symmetrically around BLO components $\omega_{L1}$ and $\omega_{L2}$. (b) Squeezing spectrum and probe gain. The red line at 0dB is the shot noise level.}
\label{fig:RF_scan}
\end{figure}
The 4WM process operates close to zero two-photon detuning and therefore generates correlations between sidebands symmetrically placed at $\omega_{HF}$ about the pump frequency, inside a bandwidth $\sigma$, as discussed in Sec.\ref{sec:introduction}. The detection of these correlations is done at a finite analysis frequency of $\omega_A = 2\pi\times1\mathrm{MHz}$, away from the 1/f technical noise band. In this case, the measurement at $\omega_A\neq 0$ cannot distinguish between positive and negative frequencies and the complete treatment requires inclusion of image bands symmetrically around the components of the BLO [see Fig.\ref{fig:RF_scan}(a)]. The measured field is then the vector sum of the fields at frequencies $\pm\omega_A$ for probe and conjugate components of the BLO and their fluctuations add independently (in quadrature). In our case the squeezing bandwidth is 2$\pi\times$40~MHz, much larger than $\omega_A$ as shown in Fig.~\ref{fig:RF_scan}(b). In these conditions both contributions $\omega_A$ and $-\omega_A$ of the photocurrent noise are a good representation of the noise at DC, that is to say the squeezing at the BLO components frequencies. Careful measurement to determine the squeezing at DC has been done in \cite{corzo2013}.

Figure \ref{fig:RF_scan}(b) shows the squeezing spectrum, obtained by scanning the frequency bands of the BLO, and the corresponding effective 4WM gain. The latter is the amplification factor of a probe seed propagating through the nonlinear medium. At high $\delta$, the amount of squeezing mirrors the gain as expected from quantum amplification theory \cite{loudon_book}. At low $\delta$, excess noise is observed as the effective 4WM gain peaks. This is due to the resonant gain being made of a large pure gain compounded with large absorption \cite{turnbull}, which results in a loss of quantum correlations between the probe and conjugate bands. Note that a similar effect could result from a lack of alignment at the $\mathrm{BS_{M}}$ for the signal and/or the BLO because of propagation effects such as cross-Kerr effect close to resonance. As the system is fundamentally multi-spatial-mode \cite{boyer}, misalignment on the $\mathrm{BS_{M}}$ mixes in uncorrelated noisy modes, resulting in measured excess noise. We have checked that the overlap remains constant by comparing the quadrature squeezing spectrum of Fig.\ref{fig:RF_scan}(b) with the intensity-difference squeezing spectrum \cite{mccormick}, which does not depend on alignment quality. The conclusion is that the spectrum in Fig.\ref{fig:RF_scan}(b) accurately reflects the amount of quadrature squeezing present in the system.

\begin{figure}[t]
\centering
\includegraphics[width=8cm]{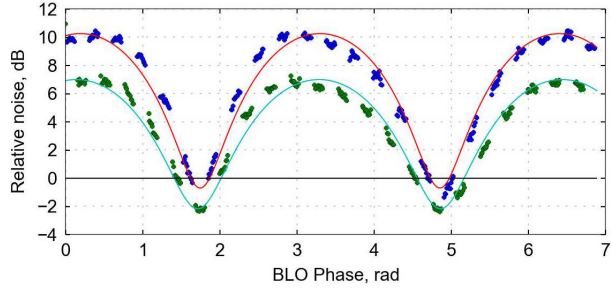}
\caption{Homodyne detector signal for two different two-photon detunings $\delta_1 = 2\pi\times12\,\mathrm{MHz}$ (green) and $\delta_2 = 0$ (blue). The measurement is taken in a single run by alternating the AOM driving frequency. The lines are cosine fits to the data and the black line represents the shot-noise-level.}
\label{fig:homodyne_signal}
\end{figure}

\subsection{Noise ellipse rotation}\label{sec:NEP_rotate}
The BLO homodyne detector operating at a small analyzing frequency, effectively DC, allows for the squeezing angle dependence on the sidebands frequency to be determined.  In other words, not only the amplitude $s(\delta)$ but also the phase $\theta(\delta)$ of the squeezing spectrum can be measured, down to an overall phase.  The measurement is performed by recording the phase of the BLO that gives the lowest noise measurement (i.e. the angle of the small axis of the noise ellipse) as the BLO detuning $\delta$ is scanned. In practice, because in our experiment the relative phases between signal components and BLO components are not stable over the long term, the homodyne measurement of the signal quadrature is performed by rapidly alternating the value of the BLO detuning, here the two-photon detuning $\delta$ of the seed beam, with a reference value $\delta_{0}=2\pi\times\mathrm{4~MHz}$, as the phase of BLO is scanned. An example of the resulting noise measurement is shown in Fig.~\ref{fig:homodyne_signal}. Besides the change in squeezing amplitude, it is obvious that the measured squeezing phase is identical for both BLO frequencies and we have checked that this is the case across the squeezing spectrum. This is to be expected since the BLO is generated in a similar fashion as the quadrature-squeezed vacuum itself. As a result, the frequency components of the BLO are subject to the same retardation effects as those for the sidebands of the signal against which they beat. 

\begin{figure}[t!]
\centering
\includegraphics[width=9cm]{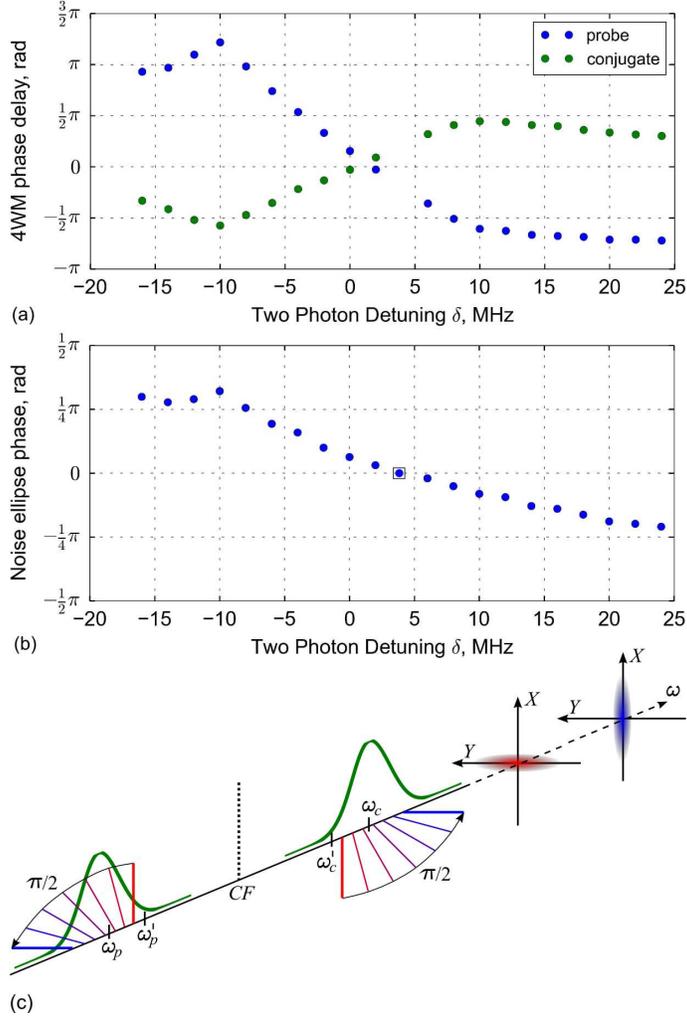}\\
\caption{Phase delay of probe and conjugate as a function of the two-photon detuning. (a) Phase delays. (b) Noise ellipse phase rotation. The phase delay is measured relatively to the delay at a detuning of $\delta_0$ (the square data point). (c) Correlations in the sideband picture. The squeezing bandwidth (green) contains the correlated sidebands, represented by sticks oriented at the angles of destructive interference of the fluctuations. When the two-photon detuning is changed from low values (red) to high values (blue) the noise ellipse rotates by $\pi/2$. The dotted line represents CF and the rotating sticks represent the rotation of the correlated sidebands resulting in rotation of the noise ellipse.}
\label{fig:NEP}
\end{figure}

Since by design the BLO is at a constant phase with respect to the signal across the squeezing spectrum, evaluating a possible noise ellipse rotation of the squeezing is here a matter of measuring the phases $\varphi_{p}^{LO}$ and $\varphi_{c}^{LO}$ of the BLO frequency components generated by the 4WM process as they scan the squeezing spectrum. This is achieved using a heterodyne beatnote technique \cite{chen} whereby the 4WM process is seeded with two beating frequencies, a fixed one as a reference and a variable one. Both components are amplified and produce conjugate components which also beat [see Fig.~\ref{fig:setup}(b)]. By recording how the phases of the resulting output amplitude beats are shifted with respect to the phase of the input beat, one can reconstruct the output phases as a function of $\delta$. Figure \ref{fig:NEP}(a) shows the result of such measurements across the squeezing spectrum, where the phases shifts have been referenced to the phase of the beat on the seed before the 4WM medium.

From Eq.~(\ref{eq:photocurrent}) one can see that the value of the phase sum $\varphi_{p}^{LO} + \varphi_{c}^{LO}$ which minimizes the noise reflects, up to a constant, the value of the squeezing angle $\theta$. Figure~\ref{fig:NEP}(b) displays the average $(\varphi_{p}^{LO} + \varphi_{c}^{LO})/2$, which represents the noise ellipse orientation, as a function of $\delta$ and shows an ellipse noise rotation of up to $\pi/2$ across the squeezing spectrum. 

\subsection{Discussion}  
Broadband squeezed light that exhibit noise ellipse rotation can be used to improve sensitivity of suspended mirror interferometers, e.g. gravitational wave detectors, beyond the standard quantum limit \cite{unruh}. Amplitude squeezing at low frequency minimizes the radiation pressure noise and phase squeezing at high frequencies minimizes the photon counting noise. In Sec.\ref{sec:NEP_rotate} we demonstrated that the SSMSS noise ellipse rotates by $\pi/2$ across the squeezing bandwidth. As shown in Fig.~\ref{fig:NEP}(c), combining this squeezed vacuum with a single bright carrier at the CF transforms it into amplitude squeezing at the low-frequency end of the squeezing spectrum and phase squeezing at the high-frequency end. Note however that like in the case of a single-frequency LO at the CF, the squeezing spectrum is centered at a frequency much higher than the squeezing bandwith, which is unsuitable for a suspended interferometer.

It is legitimate to ask whether a bichromatic bright carrier could solve this issue. Having a pair of carriers in the middle of each correlated band frequencies [$\omega_p$ and $\omega_c$ in Fig.~\ref{fig:NEP}(c)], it is always possible to choose their relative phase to ensure amplitude or phase squeezing at DC or very low analyzing frequency, as it was done in Section \ref{sec:NEP_rotate}. It is however clear from Fig.~\ref{fig:NEP}(c) that at higher analyzing frequency the noise ellipses of the contributing image pairs of correlated bands have a different orientation and collectively generate excess noise. 

It is possible to place the bright carriers on the edges of the correlated bands, at frequencies $\omega'_p$ and $\omega'_c$in Fig.~\ref{fig:NEP}(c). In this case the carrier phases can be set so that the correlated sidebands induce amplitude and phase squeezing at low and high analyzing frequencies respectively. As usual image sidebands will also contribute to the signal, but since they lie outside the squeezing spectrum in a region where the 4WM gain is unity, they will contribute half the shot noise, limiting the measurable amount of squeezing to 3~dB across the squeezing spectrum \cite{marino}. Note that these limitations do not apply to the configuration for which the 4WM process is made to operate as a phase-sensitive amplifier, as demonstrated explicitly by Corzo et. al \cite{corzo2013}.

\section{Conclusion}
We have investigated the BLO homodyne detection scheme and shown its indispensability to analyze bichromatic SSMSS where the correlated bands are separated by a frequency range inaccessible to low-noise electronics and detectors. We have measured the squeezing bandwidth of the SSMSS generated via non-degenerate 4WM to be of the order of $2\pi\times$40MHz and shown that in this bandwidth the squeezed state noise ellipse rotates by  about $\pi/2$, which is equivalent to swapping initial amplitude squeezing to phase squeezing. 

Although the band separation was 6~GHz, we expect the method to be applicable to any separation provided phase stability can be ensured between the frequency component of the BLO and the signal bands.

\section*{Funding}
This research was supported by the Engineering and Physical Sciences Research Council Grants No. EP/I001743/1 and No. EP/M013294/1. J.H. supported by the Defense Science and Technology Laboratory research PhD program via contract No. DSTL-1000092268.


\begin{thebibliography}{99}
\bibitem{footnote}
In most of the practical realizations the TMSS correlations are monochromatic, which allows for almost perfect overlap on the 50/50 beamsplitter, whereby achieving SSMSS generation independent of the propagation phases.
\end{thebibliography}
\end{document}